\documentclass[prl,floatfix,twocolumn,showpacs,amsmath,amssymb]{revtex4}
\usepackage{graphicx}
\usepackage{dcolumn}
\usepackage{bm}
\usepackage{epsfig,psfrag,amsmath,amssymb,float}
\input{epsf}
\begin{document}
\title{Superconducting gap for a two-leg tJ ladder}

\author{Didier~Poilblanc}

\affiliation{Laboratoire de Physique Th\'eorique CNRS-FRE2603,
Universit\'e Paul Sabatier, F-31062 Toulouse, France}

\author{D.~J.~Scalapino}

\affiliation{Physics department, University of California, Santa
Barbara, CA 93106}

\author{Sylvain~Capponi}

\affiliation{Department of Physics, Stanford University, Stanford,
CA 94305 \& Laboratoire de Physique Th\'eorique CNRS-FRE2603,
Universit\'e Paul Sabatier, F-31062 Toulouse, France}

\date{\today}

\begin{abstract}

Single-particle diagonal and off-diagonal Green's functions of
a 2-leg t-J ladder at $1/8$-doping are investigated by Exact
Diagonalisations techniques. A numerically tractable expression for
the superconducting gap is proposed and the frequency dependence
of the real and imaginary parts of the gap are determined. The
role of the low-energy gapped spin
modes whose energies are computed by a (one-step) Contractor
Renormalization procedure are discussed.

\end{abstract}

\maketitle

In the BCS theory of superconductivity \cite{schrieffer} the frequency
and momentum dependence of the superconducting gap provide information
about the frequency
and momentum dependence of the pairing interaction. In one spatial
dimension quantum phase fluctuations destroy long-ranged SC order
although the spectral gap is expected to survive~\cite{V98}. In a
doped two-leg spin ladder~\cite{DR96} the doped holes can form
mobile singlet pairs \cite{DRS92}
leading to dominant {\it algebraic} SC
correlations~\cite{HPNSH95} and a robust spin gap. In this Letter,
motivated by previous work by two of the authors~\cite{PS2002}, we
introduce a numerically tractable expression for the
superconducting gap function which, as we argue, contains a
non-trivial momentum
and frequency dependence apart from an overall (vanishing in the
thermodynamic limit) prefactor which accounts for 1D quantum phase
fluctuations.

We shall consider here a generic 2-leg t-J ladder,
\begin{eqnarray}
\label{ham}
&{\cal H}&=J_{\rm leg}\sum_{i,a} \vec{S}_{i,a} \cdot \vec{S}_{i+1,a}
+J_{\rm rung} \sum_{i} \vec{S}_{i,1} \cdot \vec{S}_{i,2} \\
&+& t_{\rm leg} \sum_{i,a} c_{i,a}^\dagger c_{i+1,a}
+t_{\rm rung} \sum_{i,a} c_{i,a}^\dagger c_{i,a+1}+ H.C.\, ,
\nonumber
\end{eqnarray}
\noindent where $c_{i,a}$ are projected hole operators and $a$
(=$1,2$) labels the two legs of the ladder. Isotropic couplings,
$t_{\rm leg}=t_{\rm rung}=t$ and $J_{\rm leg}=J_{\rm rung}=J$,
will be of interest here. A value like $J=0.4$ ($t$ is set to $1$)
and a doping of $\delta=1/8$ are typical of superconducting ladder
materials such as Sr$_{14-x}$Ca$_x$Cu$_{24}$O$_{41}$ and these
parameters will be assumed hereafter. Single-particle diagonal and
off-diagonal spectral functions are computed by Exact
Diagonalisations of $2\times L$ periodic ladders of size $L=12$.
The expression for the superconducting pairing function introduced
previously in the context of the two-dimensional tJ
model~\cite{PS2002} is computed numerically from the knowledge of
the diagonal and off-diagonal Green's functions of the doped 2-leg
ladder. The role of the low-energy collective triplet modes whose
energies are obtained by a
Contrator-Renormalisation~\cite{weinstein96,CP2002} (CORE)
calculation is investigated.

\begin{figure}
\begin{center}
\epsfig{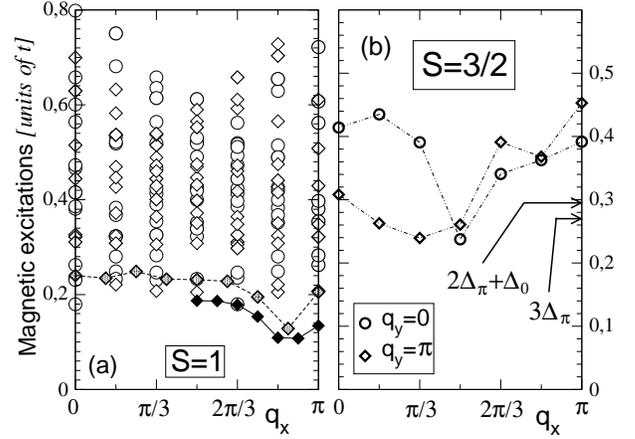} \caption{ Low-energy
magnetic excitations at $1/8$-doping and $J=0.4$ for $q_y=0$
(circles) and $q_y=\pi$ (diamonds) momenta. (a) Triplet collective
mode obtained by CORE on $2\times 16$ (shaded symbols) and
$2\times 24$ (closed symbols) ladders. The particle-hole continuum
(open symbols) constructed from the data of Figs.~\ref{fig:Akw} \&
\ref{fig:dispersion} is also shown; (b) $S=3/2$ excitations
(measured from the chemical potential) calculated by ED of the
$2\times 12$ cluster. The onsets of the 3-quasi-particle continua
are shown by arrows.} \label{fig:magnons}
\end{center}
\end{figure}

As established by numerical or Bosonisation techniques, in the
parameter range considered here, the 2-leg doped spin ladder
exhibits dominant superconducting fluctuations~\cite{HPNSH95} and
its low-energy physics is governed by two (weakly gapped)
collective spin modes (magnons), a gapped ($q_y=\pi$) charge mode
and a ($q_y=0$) zero-energy collective charge
mode~\cite{LE95,BF96,note_mode} characteristic of a C1S0 phase of
the Luther-Emery (LE) liquid universality class~\cite{LE96}. Prior
to the investigation of spectral properties we have computed the
lowest magnon excitations at hole density $1/8$ on ladders with up
to size $2\times 24$ with $N_h=6$ holes using an effective CORE
hamiltonian~\cite{CP2002} as shown on Fig.~\ref{fig:magnons}(a).
Our extrapolation gives a spin gap (for $q_y=\pi$) $\Delta_{\rm
mag}^\pi\simeq 0.11$ significantly smaller than the spin gap $\sim
J/2=0.2$ of the undoped spin ladder. On the other hand, the lowest
$q_y=0$ triplet (not shown) occurs at $\Delta_{\rm mag}^0\simeq
0.17$, close to the onset of the particle-hole continuum
(discussed below).

\begin{figure}
\begin{center}
\epsfig{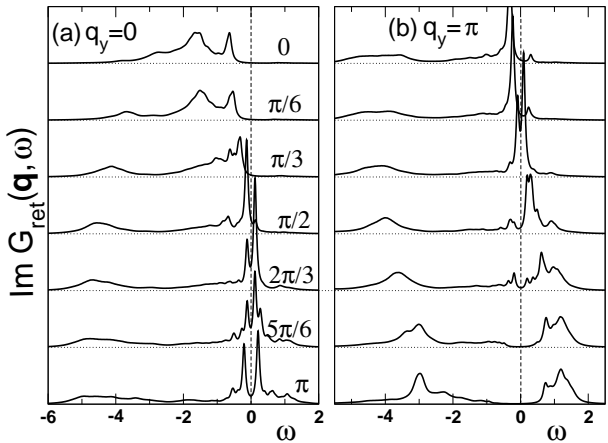} \caption{Spectral
function vs $\omega$ ($t=1$) on a $2\times 12$ ladder at
$1/8$-doping and $J=0.4$ for two transverse momenta $0$ and $\pi$
in panels (a) and (b) respectively. Data for different chain
momenta are shifted w.r.t each other.} \label{fig:Akw}
\end{center}
\end{figure}

We now turn to the investigation of the single-particle
excitations. For a finite size system, it is convenient to define
Green's functions so that the sets of electron-like (for
$\omega<0$) and hole-like (for $\omega>0$) poles coincide. The
time-ordered diagonal Green's function is then defined as,
\begin{eqnarray}
&&G({\bf q},\omega)=\langle N |{c}_{{\bf q},\sigma}^\dagger
\frac{1}{\omega-i\epsilon+H-{\bar E}_{N-1}} {c}_{{\bf q},\sigma} | N
\rangle \nonumber \\
&+&\langle N-2 |{c}_{{\bf q},\sigma} \frac{1}{\omega+i\epsilon-H+{\bar
E}_{N-1}} {c}_{{\bf q},\sigma}^\dagger | N-2 \rangle \, ,
\label{eq:gk_of_w}
\end{eqnarray}
and the retarded one, $G_{\rm ret}({\bf q},\omega)$, by the
substitution $\omega+i\epsilon\rightarrow\omega-i\epsilon$ in the
second term. $\epsilon$ is a small imaginary part set to $0.05$
hereafter. Note that the "symmetrization" between $\omega$ and
$-\omega$ implies using two ground states (GS) with particle
numbers $N$ and $N-2$ surrounding the value $N-1=N_s-N_h$
corresponding to the actual hole density ie here $N_h=3$ holes on
$N_s=2L=24$ sites. For convenience, the energy reference ${\bar
E}_{N-1}$ is defined as the average between the GS energies of $|
N \big>$ and $| N-2 \big>$. $G({\bf q}, \omega)$ and $G_{\rm
ret}({\bf q},\omega)$ can be calculated using a standard
continued-fraction method.

\begin{figure}
\begin{center}
\epsfig{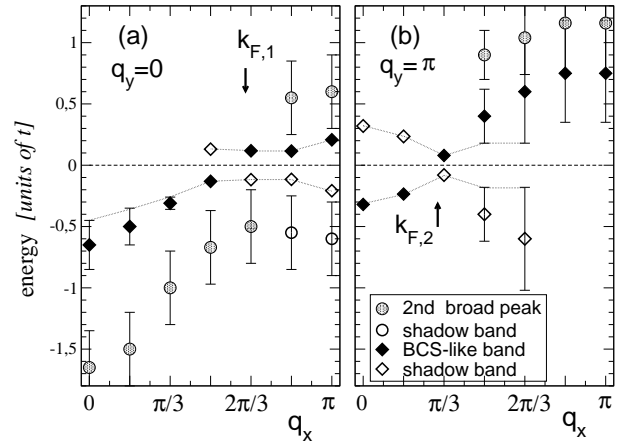} \caption{Momentum dispersion
of the low-energy peaks seen in Fig.~\protect\ref{fig:Akw}. Error
bars give the widths of the peaks (if any) and the dotted lines
correspond to the lowest excitation energies.}
\label{fig:dispersion}
\end{center}
\end{figure}

The data for the single-particle spectral function $A({\bf
q},\omega)=-\frac{1}{\pi}\rm{Im}G_{\rm ret}({\bf q},\omega)$ shown
in Fig.~\ref{fig:Akw} can be fairly well described by (i) BCS-like
$q_y=0$ (bonding) and $q_y=\pi$ (antibonding) quasi-particle bands
with the chemical potential reference energy $\omega=0$ and (ii) a
broad incoherent background extending further away towards
negative energies in agreement with a calculation on a smaller
$2\times 8$ cluster at the same hole
density~\cite{spectral_ladder}. Note that the peaks of the
low-energy band-like features narrow near the chemical potential.
Note also that the approximate Fermi momenta $(k_{F,1},0)$ which
lies between $(5\pi/6, 0)$ and $(2\pi/3,0)$ and $(k_{F,2},\pi)\sim
(\pi/3,\pi)$ are in rough agreement with Luttinger's
theorem~\cite{GHR98} (which gives $k_{F1} + k_{F2}=7\pi/8$). The
low-energy peaks of $A({\bf q},\omega)$ plotted vs momentum in
Fig.~\ref{fig:dispersion} exhibit BCS-like dispersions
$\pm\sqrt{\tilde \epsilon^2_{(q)} +\Delta^2(q)}$ with the
magnitudes of the gaps $\Delta (k_{F1}, 0)\equiv\Delta_0\simeq
0.115$ and $\Delta (k_{F2},\pi)\equiv\Delta_\pi\simeq 0.090$.
Here, the solid points denote the maximum spectral weight and the
open circles the BCS-like shadow band. Note that the particle-hole
continuum can been obtained by considering all combinations of any
two of the lowest single-particle excitations.
Figs.~\ref{fig:magnons}(a) and \ref{fig:magnons}(b) give further
evidence that bound triplet-hole pair excitations sits below this
continuum~\cite{LE95}.

\begin{figure}
\begin{center}
\epsfig{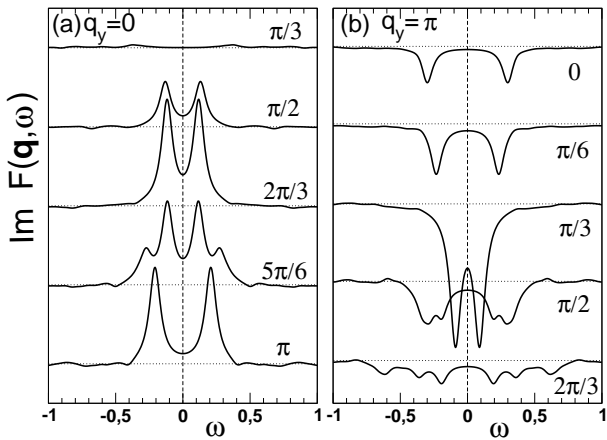} \caption{Spectral weight
of the superconducting Green's function. Same conventions as in
Fig.~\protect\ref{fig:Akw}. Note the change of sign between
transverse momenta $0$ and $\pi$. } \label{fig:Fkw}
\end{center}
\end{figure}

\begin{figure}
\begin{center}
\epsfig{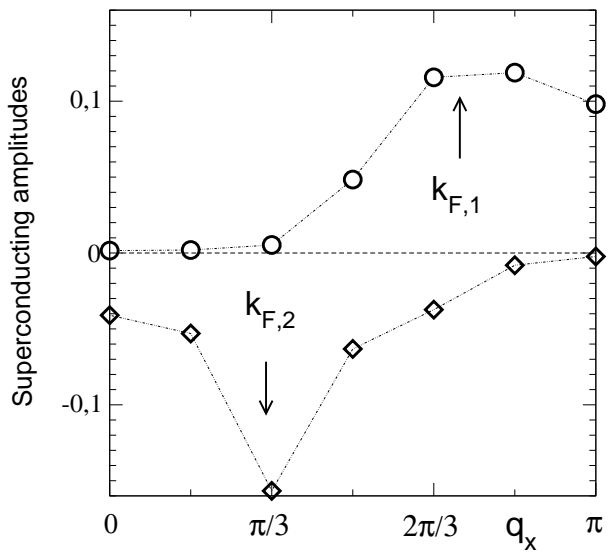} \caption{
Superconducting amplitude as a function of the chain momentum
$q_x$ for the bonding $q_y=0$ (solid circles) and antibonding
$q_y=\pi$ (open circles) transfer momenta.} \label{fig:super}
\end{center}
\end{figure}

The superconducting Gorkov's off-diagonal one-electron
time-ordered Green's function~\cite{schrieffer} defined by
\begin{equation}
F({\bf q},t)=i\langle T {c}_{{\bf -q},-\sigma}(t/2) {c}_{{\bf
q},\sigma}(-t/2)\rangle\, , \label{eq:fk_of_tau}
\end{equation}
can be computed in a finite system by taking the expectation value
between the two GS $| N \big>$ and $| N-2 \big>$ differing by two
particles, hence reflecting superconducting fluctuations. Note
that these states are both spin singlets so that the (Fourier
transformed in time) Green-function, $F({\bf q},\omega)$, is {\it
even} in frequency and can be expressed as,
\begin{equation}
F({\bf q},\omega)={\tilde F}_{\bf q}(\omega+i\epsilon) +{\tilde
F}_{\bf q}(-\omega+i\epsilon)\, , \label{eq:fk_of_w}
\end{equation}
where
\begin{equation}
{\tilde F}_{\bf q}(z)=\langle N-2 |{c}_{{\bf -q},-\sigma}
\frac{1}{z-H+{\bar E}_{N-1}} {c}_{{\bf q},\sigma} | N
\rangle \, , \label{eq:fk_of_z}
\end{equation}
is defined for all complex $z$ (with ${\rm Im}\,z\ne 0$).
Following Ohta, et.~al \cite{OSEM94} who extended the
continued-fraction method to deal with off-diagonal Green's
functions, we have computed $F({\bf q},\omega)$ on a $2\times 12$
ladder at a hole density of $1/8$ and for all momenta. Sharp low
energies features are found near momenta $(k_{F,1},0)$ and
$(k_{F,2},\pi)$ as shown in Fig.~\ref{fig:Fkw}. As the diagonal
Green's function, the off-diagonal one is also gapped. The
equal-time pair amplitude $\big< c_{-{\bf q},-\sigma} c_{{\bf
q},\sigma} \big>$ corresponding to the integrated weight
$\frac{1}{2i\pi} \int_{-\infty}^\infty F({\bf q},\omega)\,
d\omega$ is plotted in Fig.~\ref{fig:super} showing maxima in
the vicinity of the (estimated) Fermi momenta. Note also in
Figs.~\ref{fig:Fkw} and \ref{fig:super} that the pair
amplitudes have opposite signs for the two transverse momenta $0$
and $\pi$, reminiscent of a two-dimensional (2D) d-wave orbital
symmetry.

A long-ranged superconducting GS is characterized by a
frequency-dependent gap function $\Delta_{\rm SC}({\bf q},\omega)$
directly proportional to $F({\bf q},\omega)$.
A definition such as
\begin{equation}
\Delta_{\rm SC}({\bf q},\omega)= \frac{2\omega F({\bf q},\omega)}{G({\bf
q},\omega)-G({\bf q},-\omega)} \, ,\label{eq:delta}
\end{equation}
is consistent with Nambu-Eliashberg theory~\cite{schrieffer} as
shown by two of the authors~\cite{PS2002}. From a numerical
calculation of $G({\bf q}, \omega)$ and $F({\bf q}, \omega)$,
$\Delta_{\rm SC}({\bf q},\omega)$ has been obtained in a 2D t-J
model~\cite{PS2002}, extending previous work by Ohta
et.~al~\cite{OSEM94} who fit the spectral weight ${\rm Im} F({\bf
q}, \omega)/\pi$ to a d$_{x^2-y^2}$ BCS-Bogoliubov quasi-particle
form~\cite{schrieffer}. In contrast to a true SC state, a doped
2-leg ladder (for parameters used here) exhibits long distance
power law pair field correlations which decay as $x^{-1/K\rho}$.
Here $K_\rho$ is the Luttinger liquid parameter associated with
the massless charge mode. This implies~\cite{bozcalc} that for a
ladder of length $L$, the off-diagonal Green's function $F({\bf
q},\omega)$ decays as $(\xi/L)^{1/2K_\rho}$.  Here the coherence
length $\xi$ is proportional to the inverse of the spin gap. Thus,
we expect that $\Delta({\bf q},\omega)$ given by
Eq.~(\ref{eq:delta}) will vary as $(\xi/L)^{1/2K_\rho}$. Using a
CORE calculation supplemented by conformal invariance
identities~\cite{CP2002}, we obtain, for $\delta=1/8$ and $J=0.4$,
$K_\rho \simeq 0.65$ in agreement with previous ED
evaluations~\cite{LE96} and DMRG data~\cite{siller01}. We then
expect the SC gap function $\Delta_{\rm SC}({\bf q},\omega)$ to
vanish in the thermodynamic limit due to Cooper pair phase
fluctuations. Indeed, using e.g. a standard low-energy
long-wavelength LE field theory, it can be shown that the SC
Green's function decays with system size like
$(1/L)^\frac{1}{2K_\rho}$ due to SC phase
fluctuations~\cite{bozcalc}. However, apart from this prefactor,
$\Delta_{\rm SC}({\bf q},\omega)$ {\it calculated on a finite
system} can provide informations on the dynamics of pairing at
intermediate distances.

\begin{figure}
\begin{center}
\epsfig{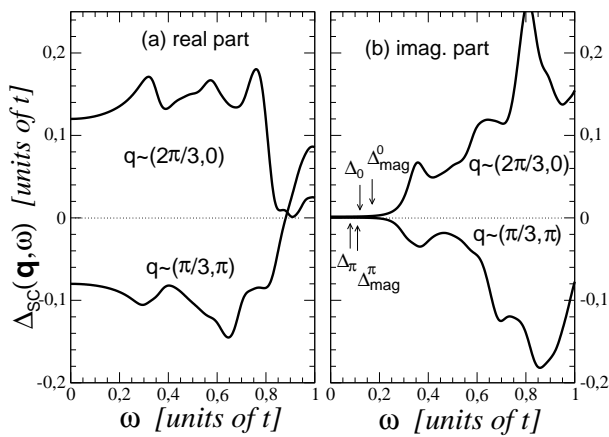} \caption{Real (a) and
imaginary (b) parts of the superconducting gap vs $\omega$,
normalization as discussed in the text,
obtained by ED results for a $2\times 12$ t-J ladder at the two
bonding and antibonding Fermi momenta $(2\pi/3, 0)$ and $(\pi/3, \pi)$.
An imaginary damping $\epsilon=0.05$ is used and the irrelevant offset
$F(0,0) \propto i\epsilon$ has been subtracted.}
\label{fig:super2}
\end{center}
\end{figure}

In our case $L/\xi$ is only of order 2 to 3 so that the scaling
factor $(\xi/L)^{1/2K_\rho}$ is of order one and we will simply
normalize \cite{gapdef} $\Delta_{SC}({\bf q}, \omega)$ so that
$\Delta_{SC} (k_{F1}, 0, \omega=0)=0.12t$ and $\Delta_{SC}
(k_{F2}, \pi, \omega=0)=-0.09t$.  Using this normalization, we
have plotted the real and imaginary parts of $\Delta_{SC}({\bf q},
\omega)$ for the bonding ${\bf q}=(2\pi/3, 0)$ and antibonding
${\bf q}=(\pi/3, \pi)$ Fermi momenta in Fig.~\ref{fig:super2}. As
expected for a d-wave-like gap, the sign of the gap changes when
one goes from $(k_{F1}, 0)$ to $(k_{F2}, \pi)$.  Otherwise, the
frequency dependence of both the real and imaginary parts of the
gap is quite similar.  The imaginary part of the gap appears to
onset at values of $\omega \sim \Delta_0 + \Delta^\pi_{\rm mag}$
where $\Delta_0\sim 0.12$ is the superconducting gap and
$\Delta^\pi_{\rm mag}\sim 0.11$ is the magnon gap.  The imaginary
part of the gap then increases until one passes through the
particle-hole spectrum \cite{LE95} and then decreases at yet
higher energies. The real part of the gap increases slightly and
then when $\omega$ increases beyond the electron-hole spectrum,
the real part of the gap drops.

Thus, we believe that this approach to calculating
$\Delta_{SC}({\bf q}, \omega)$ allows one to probe the internal
structure of a pair. Clearly, additional calculations,
particularly for the 2-leg Hubbard ladder will be of interest in
providing further insight into the relationship between frequency
dependence of the gap and the dynamics of the pairing interaction.

\noindent{\bf Acknowledgments}

D.J.~Scalapino would like to acknowledge support from the
US Department of Energy under Grant No DE-FG03-85ER45197.
D.~Poilblanc thanks M.~Sigrist (ETH-Z\"urich) for
discussions and aknowledges hospitality of the Physics Department
(UC Santa Barbara) where part of this work was carried out.
Numerical computations were done on the vector NEC-SX5 supercomputer
at IDRIS (Paris, France).
We thank E.~Orignac for useful comments and IDRIS (Orsay, France)
for CPU-time on the NEC-SX5 supercomputer.


\begin{thebibliography}{99}

\bibitem{schrieffer} J.R.~Schrieffer, {\it Theory of
Superconductivity} (Benjamin, NY 1964).

\bibitem{V98} J.~Voit, {\sl Eur.~Phys.~J.~B} {\bf 5}, 505 (1998);
F.H.L.~Essler and A.M.~Tsvelik, cond-mat/0205294 (2002).

\bibitem{DR96} E.~Dagotto and T.M.~Rice, {\sl Science} {\bf 271}, 618
(1996).

\bibitem{DRS92} E.~Dagotto, J.~Riera, and D.J.~Scalapino, {\sl Phys.~Rev.~B}
{\bf 45}, 5744 (1992).

\bibitem{HPNSH95} C.~A.~Hayward et al., {\sl Phys.~Rev.~Lett.} {\bf 75},
926 (1995).

\bibitem{PS2002} D.~Poilblanc and D.J.~Scalapino, {\sl Phys.~Rev.~B} {\bf 66},
052513 (2002).

\bibitem{weinstein96} C. J. Morningstar and M. Weinstein, {\sl Phys.
Rev. D} {\bf 54}, 4131 (1996). For Hubbard models see E.~Altman
and A.~Auerbach, {\sl Phys.~Rev.~B} {\bf 65}, 104508 (2002).

\bibitem{CP2002} On the $2\times 24$ ladder,
$2\times 2$ plaquette bosonic triplet and hole pair
states are retained to construct a range 2 effective hamiltonian.
On the $2\times 16$ ladder, additional one-hole fermionic plaquette
states are also kept;
S.~Capponi and D.~Poilblanc, {\sl Phys.~Rev.~B} {\bf 66},
180503 (2002).


\bibitem{LE95} D.~Poilblanc, D.J.~Scalapino, and
W.~Hanke, {\sl Phys.~Rev.~B} {\bf 52}, 6796 (1995); D.~Poilblanc
et al., {\sl Phys.~Rev.~B} {\bf 62}, R14633 (2000).

\bibitem{BF96} L.~Balents and M.P.A.~Fisher, {\sl Phys.~Rev.~B}
{\bf 53},  12133 (1996) and references therein.

\bibitem{note_mode} The second charge mode is gapped.

\bibitem{LE96}
M.~Troyer, H.~Tsunetsugu, and T.M.~Rice, {\sl Phys. Rev. B} {\bf
53}, 251 (1996); C.A.~Hayward and D.~Poilblanc, {\sl Phys.~Rev.~B}
{\bf 53}, 11721 (1996).


\bibitem{spectral_ladder} D.~Poilblanc, J.~Riera and E.~Dagotto,
{\sl Eur.~Phys.~J.~B} {\bf 7}, 53 (1999).

\bibitem{GHR98} P.~Gagliardini, S.~Haas, and T.~M.~Rice,
{\sl Phys. Rev. B} {\bf 58}, 9603 (1998).


\bibitem{OSEM94} Y.~Ohta,
T.~Shimozato, R.~Eder, and S.~Maekawa, {\sl Phys.~Rev.~Lett.} {\bf
73}, 324 (1994).

\bibitem{bozcalc} A bozonization calculation shows that $\Delta(q,\omega)$
given by Eq.~(\ref{eq:delta}) factors into an $L^{-1/2K_\rho}$
factor times a function of $\omega$ and $q$ which is independent
of L. E.~Orignac and D.~Poilblanc, cond-mat/0303053.

\bibitem{siller01} T.~Siller, M.~Troyer, T.M. Rice and S.R. White,
{\sl Phys.~Rev.~B} {\bf 63}, 195106 (2001).

\bibitem{gapdef} Strickly speaking, the gap should be defined at the gap edge
so that, for example $\Delta_{SC} (k_{F1}, 0,\omega = \Delta_0
(k_{F1}, 0)) = \Delta_0 (k_{F1}, 0)$.  However, since $\Delta_{SC}
({\bf q}, \omega)$ is relatively flat out to energies of order
$\Delta_0$, we have set $\omega=0$.


\end{thebibliography}
\end{document}